\documentclass[referee]{raa}    

\usepackage{graphicx,times}             
\usepackage{natbib}
\usepackage{amssymb,amsmath}
\bibpunct{(}{)}{;}{a}{}{,}
\usepackage[a4paper=true,dvipdfm=true,pagebackref=true]{hyperref}
\hypersetup{colorlinks = true, linkcolor = green, anchorcolor = red, citecolor = blue, filecolor = red, pagecolor = red, urlcolor = red}
\begin{document}

   \title{Multi-wavelength observations of the 2014 June 11 M3.9 flare: temporal and spatial characteristics
 }
   \volnopage{Vol.0 (20xx) No.0, 000--000}      
   \setcounter{page}{1}          

   \author{Damian J. Christian
      \inst{1}
   \and David Kuridze
      \inst{2,3}
   \and David B. Jess
      \inst{3}
    \and Menoa Yousefi
      \inst{1}
    \and Mihalis Mathioudakis
      \inst{3}
   }

   \institute{Department of Physics and Astronomy, California State University Northridge, Northridge, CA 91330, USA; {\it damian.christian@csun.edu}\\
        \and
       Institute of Mathematics, Physics and Computer Science, Aberystwyth University, Ceredigion, Cymru, SY23 3, UK\\
        \and
         Astrophysics Research Center, School of Mathematics and Physics, Queen\'s University Belfast, Belfast BT7 INN, UK\\
\vs\no
   {\small Received~~2018 month day; accepted~~20xx~~month day}}

\abstract{We present multi-wavelength observations of an M-class flare (M3.9) that occurred on 2014 June 11. Our observations were conducted with the Dunn Solar Telescope (DST), adaptive optics, the multi-camera system ROSA (Rapid Oscillations in Solar Atmosphere) and new HARDcam (Hydrogen-Alpha Rapid Dynamics) camera in various wavelengths, such as Ca~II~K, Mg~I~b$_2$ (at 5172.7 \AA), and H$\alpha$ narrow-band, and G-band continuum filters.  Images were re-constructed using the Kiepencheuer-Institut Speckle Interferometry Package (KISIP) code, to improve our image resolution. We observed intensity increases of $\approx$120-150\% in the Mg, Ca~K and H$\alpha$ narrow band filters during the flare. Intensity increases for the flare observed in the SDO EUV channels were several times larger, and the GOES X-rays increased over a factor of 30 for the harder band. Only a modest delay is found between the onset of flare ribbons of a nearby sympathetic flare and the main flare ribbons observed in these narrow-band filters. The peak flare emission occurs within a few seconds for the Ca~K, Mg, and H$\alpha$ bands. Time-distance techniques find propagation velocities of $\approx$60 km/s for the main flare ribbon and as high as 300 km/s for smaller regions we attribute to filament eruptions. This result and delays and velocities observed with SDO ($\approx$100 km/s) for different coronal heights agree well with the simple model of energy propagation versus height, although a more detailed model for the flaring solar atmosphere is needed. And finally, we detected marginal quasi-periodic pulsations (QPPs) in the 40--60 second range for the Ca~K, Mg and H$\alpha$ bands, and such measurements are important for disentangling the detailed flare-physics.
} 

   \authorrunning{D.~J. Christian et al.}            
   \titlerunning{Multi-wavelength observations of a M3.9 flare}

   \maketitle
%
%
\section{Introduction}           
\label{sect:intro}

Solar flares vary in magnitude and frequency from rare, large, X-class flares, to common micro-flares and other sub-arcsecond explosive events \citep{J18}.  In flares, a rapid energy transfer occurs between the corona, chromosphere, and photosphere through non-thermal electron beams, radiation, conduction, Magnetoacoustic/Alfv\'en waves, and mass motions. The intermittent nature of the non-thermal electron beams can result in very rapid variations in chromospheric and coronal emission. These variations arise from a combination of energy/ionization imbalance and chromospheric condensation, and are determined by the intensity of the non-thermal energy flux deposited in the lower atmosphere. The chromospheric plasma is heated to very high temperatures (5--30 MK) creating an overpressure, leading to an expansion into the overlying atmosphere. This in turn results in blueshifts in upper chromosphere and transition region lines, a process known as chromospheric evaporation \citep{M06, P07, M15}. On the other hand, due to the upward and downward momentum balance there should be down flowing pattern (condensation) as a back reaction of chromospheric evaporation, leading a redshifts of tens of kilometers per second in H$\alpha$ \citep{K15}.
Additionally, many flares are accompanied by white light emission. White light flare emission is observed in the near-UV and optical continuum, often appearing in transient footpoint regions 
\citep{N89, I07,J08}, and see reviews by \citet{HWM06,H11}, for example. 
The energy content of the white light component of a flare may actually exceed the soft X-ray component by a factor of 100 \citep{H11}. 

There are many open questions on the conditions in the photosphere and chromosphere pre-flare, and their response during the flare.  Recent advances in solar imaging in both spatial and temporal resolution allow us to measure the response of the photosphere and chromosphere at multiple heights in the solar atmosphere.  Thus, the morphology and temporal changes of the flare emission observed in different narrow-band filters (such as Ca K, Mg) promise to help constrain the detailed mechanisms of flare energy production and release, e.g. reviews by  \citet{F11, H11}.  \citet{L83}, using observations in the Mg I b$_2$ line, found Mg flare kernels resembled white-light flare kernels, and associated the Mg emission with the impulsive flare phase. More recently, \citet{B14} used Mg~I~b$_2$ to probe the poorly constrained chromospheric magnetic field.
\newpage
In the present paper, we employ these new capabilities of higher spatial and temporal resolution of ground-based instrumentation and report on an M-class flare observed at several layers in the solar atmosphere. We observed an M3.9 flare observed on 2014 June 11 (SOL2014-06-11T21:03 UT), with the Rapid Oscillations of the Solar Atmosphere or (ROSA), HARDcam and CSUNcam instruments at the National Solar Observatory's Dunn Solar Telescope. In \S\ 2 we present the ROSA and ground-based observations and data analysis, and inclusion of observations from NASA's Solar Dynamic Observatory (SDO). 
In \S\ 3 we present results for the flare characteristics, and compare these to previous observations and discuss the flare propagation in simple models and search for quasi-periodic pulsations (QPPs) in \S\ 4. Lastly, in \S\ 5 we summarize our findings.


\section{Observations}
\label{sect:Obs}

Our data were obtained with the Rapid Oscillations in the Solar Atmosphere \citep{J10} (ROSA; Jess et al. 2010) camera system. ROSA is a synchronized, multi-camera high cadence solar imager installed on the Dunn Solar Telescope (DST) at the National Solar Observatory, Sacramento Peak, NM.  The DST pointing was at S18.5E60.0, covering AR 12087 (at heliocentric coordinates, --755$\arcsec$, --295$\arcsec$). The observations sequence started at 19:20 UT, and continued until 21:31 UT, with a span of just over 2 hours and 11 minutes.  The ROSA continuum channels recorded 30 fps and narrow-band Mg I b$_2$ filter was obtained at half of this data rate. The ROSA set-up used with a wider field of view than many previous studies, $\approx2\arcmin$ giving a resolution of 0.12$\arcsec$ per pixel or a 2-pixel diffraction-limited resolution of $\approx$173 km.  ROSA was combined with the Hydrogen-Alpha Rapid Dynamics camera (HARDcam) \citep{JdMM12} and CSUNcam \citep{G15}, using H$\alpha$ and Ca~II~K (hereafter Ca~K) filters at a frame rate of 30 fps, respectively. Central wavelengths and filter properties for the H$\alpha$ and Ca K are given in Jess et al. 2010. Interferometric BIdimensional Spectrometer \citep{C06, RC08} observations using the Na D$_1$ line were also obtain and have been presented by \citep{KMC16}. The observations were obtained with high-order adaptive optics  \citep{R04} to correct wavefront deformations in real-time.  Although, the overall seeing was poor at times and the observations hampered by clouds, we were able to measure the flare properties and perform speckle image reconstruction. The Kiepenheuer-Institut Speckle Interferometry Package (KISIP) speckle reconstruction algorithm \citep{W08}, was used on the images, with a 64 to 1 restoration.  The Mg~I~b$_2$ (hereafter Mg) data was obtained with a cadence of 15.15 fps, and the reconstructed cadence is 4.22 seconds per image. The H$\alpha$ and Ca K data were obtained at 30.3 fps and both have a reconstructed cadence of 2.12 seconds per image. We searched for quasi-periodic fluctuations in our narrow-band data using wavelet techniques as described in \citet{J07}. Our ground-based observations caught the M3.9 flare rising at $\sim$20:53 UT and peaking near 20:57 UT, and the GOES X-ray bands measured the flare peak at $\sim$21:03 UT. We show a sequence of flare images in the Ca~K, Mg, H$\alpha$, and G-band in Figure 1. Flares light curves for all 3 bands were extracted in a 34$\arcmin \times  26\arcmin$  region. 

   \begin{figure}
   \centering
   \includegraphics[width=15cm, angle=-0]{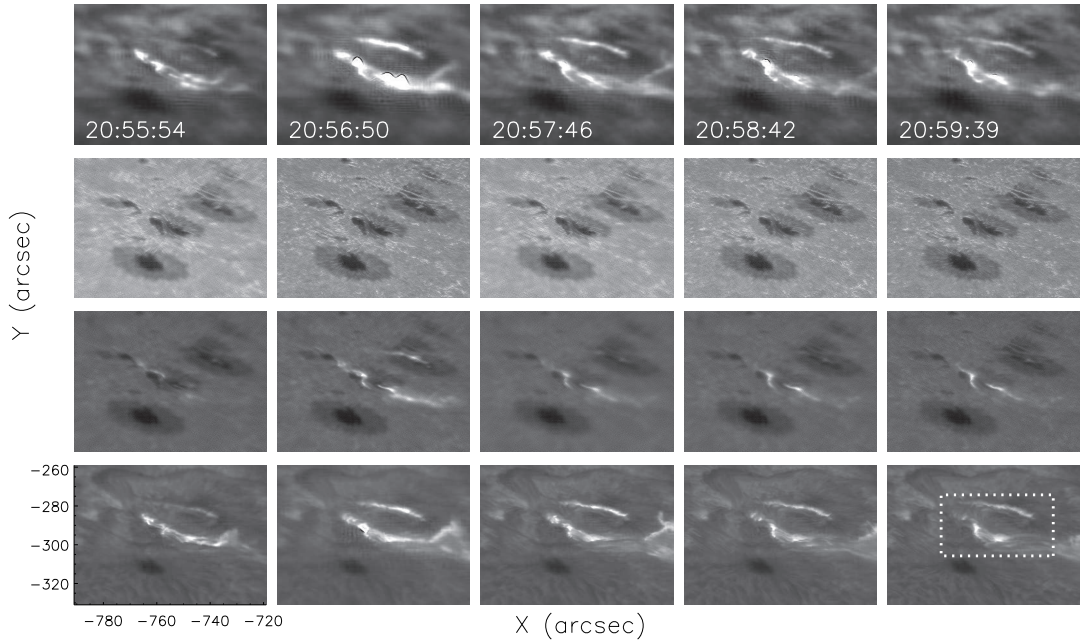}
   \caption{Image sequence from ground-based observations from ROSA, HaRDcam and CSUNcam spanning the 2014 June 11 M3.9 class flare peak at intervals of about 1 minute.
Filters from top to bottom are: Ca~K, G-band, Mg~I~b$_2$ and H$\alpha$.  The extraction region for flare light curves is indicated in the lower right panel for H$\alpha$.
}
   \label{Fig1}
   \end{figure}
   
\subsection{Solar Dynamic Observatory Observations}

We further supplemented our ROSA observations with EUV data from the Atmospheric Imaging Assembly (AIA) on-board the Solar Dynamics Observatory  \citep{L12} and magnetic information from the Helioseismic and Magnetic Imager \citep{S12, SSB12}. The AIA instrument images the entire solar disk in 10 different channels, incorporating a two-pixel spatial resolution of 1.2$\arcsec$ ($\approx$900 km for the AIA's PSF) and a cadence of 12 sec for the EUV channels and 24 sec for the 1600 and 1700 \AA channels. Here, we selected 5 EUV datasets spanning 20:30 -- 22:00 UT on 2014 June 11, consisting of 445 images in each of the 94, 131, 193, 211, 304, \& 335 \AA\ AIA channels, 
and 222 images for the 1600 and 1700 \AA\ channels.  The SDO observations caught the M3.9 flare starting at approximately 20:53 UT and the subsequent brightening peaking at near 20:57 for the cooler SDO channel and $\sim$21:06 UT for the hotter channels. Light curve for the different SDO channels were extracted in a $120\arcsec \times 90\arcsec$ region. Sample SDO images along with our Mg and H$\alpha$ bands are shown in Figure 2.  We also obtained the HMI line-of-sight (LOS) magnetograms from 20:48 and 21:12 UT, before and after the flare, respectively. The HMI images were aligned and  degraded to the AIA resolution with hmi\_prep.

   \begin{figure}
   \centering
   \includegraphics[width=15cm, angle=-0]{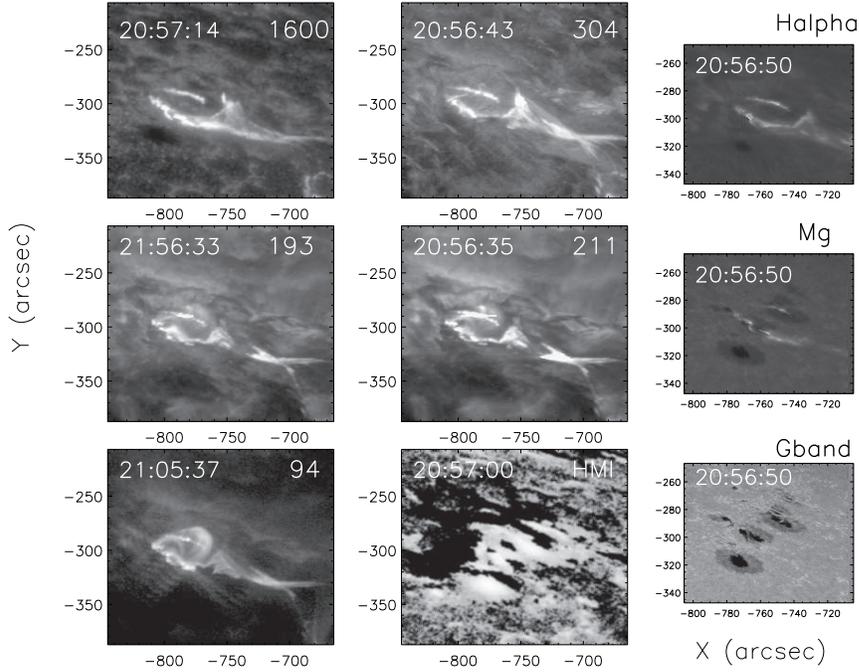}
   \caption{SDO/AIA flare images in 5 different bands near the flare peak, along with our HARDCam H$\alpha$, and ROSA G-band and Mg observations in the right column. The HMI LOS magnetogram is also shown in the lower middle panel (see text).   (Note: Images were selected closest to the time of the flare peak in each band, but without saturation.)
   }
  \label{Fig2}
   \end{figure}

\section{Results}
\subsection{Flare Properties}
Our multi-wavelength observations of AR 12087 detected an M3.9 flare on 2014 June 11. 
The GOES soft (1--8 \AA) and hard (0.5--4.0 \AA) X-rays peaked at 21:03 UT and rose factors of 15 and 33 over the quiescent emission, respectively.  The flare in the ROSA Mg~I~b$_2$ light curve rose a factor of 1.24 over the quiescent emission and peaked near 20:56:51 UT. The Ca~K and H$\alpha$ light curves increased a factor of 1.55 and 1.5 over their pre-flare, quiescent emission, respectively. In Figure 3 we compare the narrow-band filter data to that obtained from the GOES X-ray satellite. The rise time for the Mg flare was $\sim$105 sec, with a decay times of the narrow band filters exceeding 30 minutes and the overall emission stays larger than the pre-flare quiescent emission for the remaining time of our observations. The GOES peak for our M3.9 flares is $3.9\times10^{-5} W/m^2$. This works out to a total energy in the 1.0 to 8.0 \AA\ band of $1.1\times10^{29}$ ergs for the 17 minute flare duration.  We summarize the flare parameters in different bands in Table 1 extracted from the regions noted in Section 2.

   \begin{figure}
   \centering
   \includegraphics[width=12cm, angle=-0]{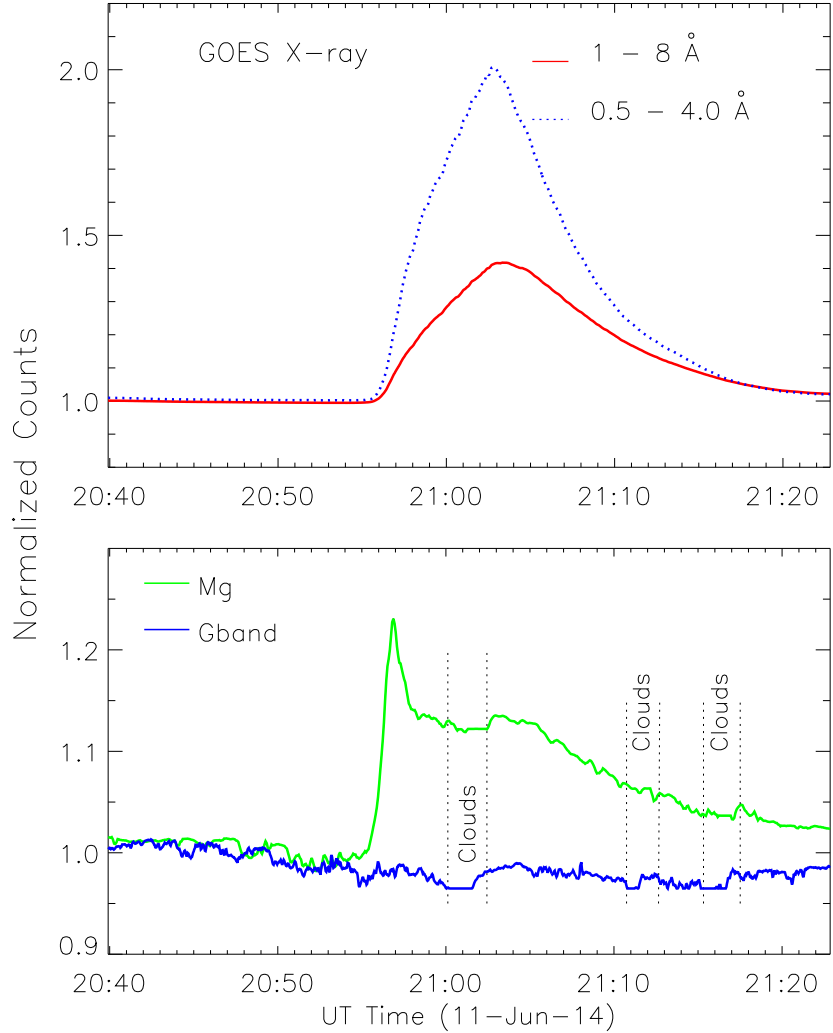}
\caption{Sample light curves for the 2014 June 11 M3.9 flare as a function UT times on 2014 June 11 . The top panel shows both the GOES soft X-ray channels with the 0.5 -- 4 \AA\ displayed in blue and the 1.0 -- 8.0~\AA\ in red.  The lower panel shows the ROSA Mg (green) and G-band (blue) light curves. Intervals disrupted by clouds are indicated in the lower panel.  
}
  \label{Fig3}
   \end{figure}

\begin{table}
\begin{center}
\caption{Light Curve Results}

 \begin{tabular}{lcccc}
   \hline\noalign{\smallskip}
      \hline\noalign{\smallskip}
$\lambda$ &	Peak   &  Rise & Decay &	Peak/Quiescent \\
                   &    (UT)	  &   (min)    &  (min)           &      \\ \hline

G-band	               & No Flare  & ... & ... & ...	 \\
Ca K         & 20:56:49 & 	$\sim$1.5	  & $>$30 & 	1.55  \\
Mg	         & 20:56:51 &	$\sim$1.7 &	$\sim$30 &	1.24  \\
H$\alpha$	&20:56:51  &	$\sim$1.5 & 	$>$35 & 	1.50  \\
SDO 304 & 20:56:43 & 	$\sim$2   & $\sim$46 	& 4.3 \\
SDO 1600	       & 20:56:31   & 	$\sim$31 	 &  $\sim$28 & 4.6  \\
SDO 193\footnotemark 	& 21:02:42  & 	$\sim$7   & $\sim$28	         & 3.4 \\
SDO 211$^1$ 	& 20:58:01  &	$\sim$2  & 	1.4 & 	1.7  \\   
SDO 335$^1$ & 20:56:51 & 	$\sim$2   & $\sim$1.5 	& 2.3 \\
SDO 94$^1$  &	21:04:25 & $\sim$9	&  $\sim$33 & 	12.7 \\
SDO 131$^1$& 21:03:32 & $\sim$8 & $\sim$29 &  22 \\
Hard X-Ray     & 	21:03   &	$\sim$10 	 & $\sim$17  & 	33  \\
Soft X-Ray      & 	21:03  &	$\sim$10 	 & $\sim$19  &	15 \\
  \noalign{\smallskip}\hline
\end{tabular}

\noindent\text{$^1$Flare peaks occurs several minutes later, consistent with "euv late phase" emission (see text).} 
  \end{center}
\end{table}

\subsection{Flare Morphology}
\label{sect:Morph}
The flare emerged in the Ca K, Mg and H$\alpha$ images as 2 separate flare ribbons that evolved into the lower longer flare ribbon (``main'' or ''lower'') with a length of about 25 Mm and a smaller sympathetic flare ribbon  (''upper'') about 18 Mm long separated by 7 to 10 Mm above.  We looked for differences in the flare properties across the flare ribbons in different regions. We show the flare light curves extracted all three bands from these different regions in Figure 4a along with the flare image for H$\alpha$. A movie of the H$\alpha$ images during the flare is also linked to Figure 4a.  There was essentially no difference in the rise times between the east and west regions surrounding the lower main flare ribbon. Moving from east to west along the lower flare ribbon find similar rise times, with the largest peak from region labeled box ``5''. Although in the main flare ribbons, the left region (box labeled ``0'') rises about 30 seconds before the emission from the upper (sympathetic) flare ribbon (labeled box ``7''). The ribbon appears to be erupting along the east-to-west direction, but not in a linear fashion that makes a propagation velocity easily calculated.  The more eastward and westward regions showed a more gradual increase than the central regions and only reached a levels $\approx$1.5 times higher than the quiescent emission, about 40\% lower than the flare peaks for the central regions of the main flare ribbon.  In the next section we investigate velocities using time-distance techniques.

 \begin{figure}[h!]
   \centering
   \includegraphics[width=12cm, angle=-0]{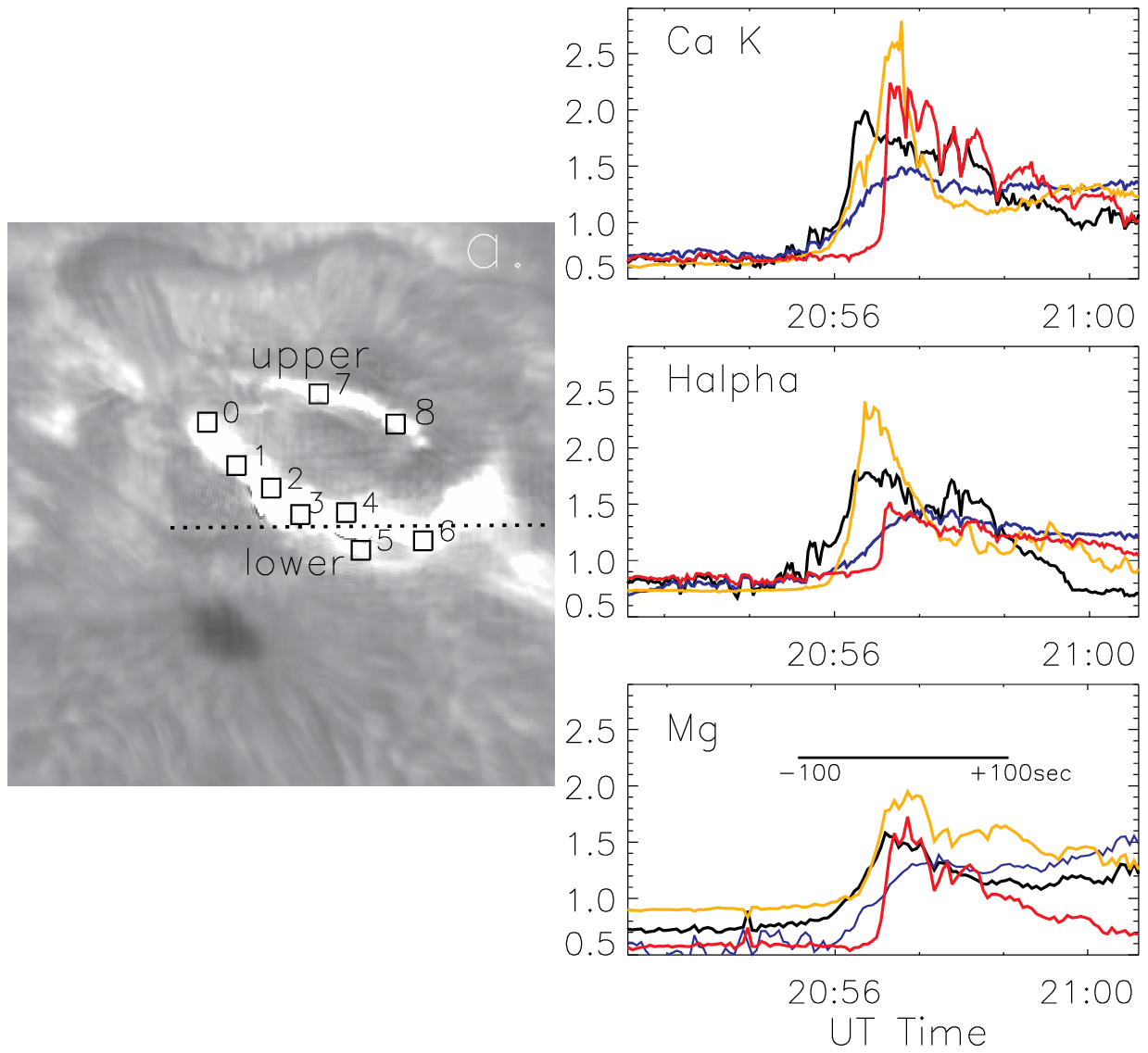}
   \vspace{-0.1in}
     \includegraphics[width=10cm, angle=-0]{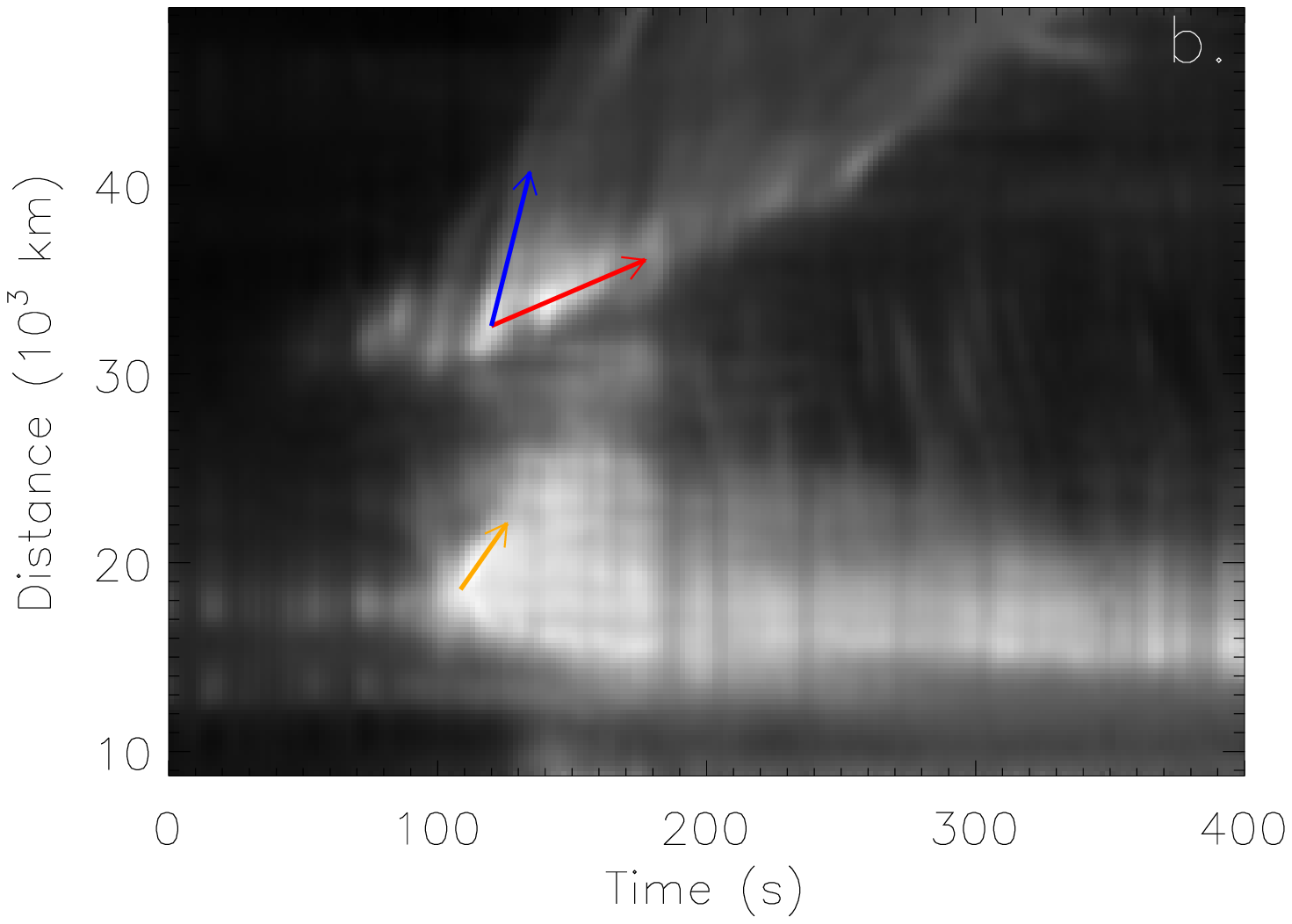}
\caption{Shown are comparisons of the H$\alpha$, Ca~K and Mg light curves extracted from different spatial regions during the M3.9 flare and a time-distance plot for H$\alpha$. a. (upper panels) The regions the light curves were extracted from are shown in the left upper panel for H$\alpha$ only. The dashed horizontal line shows where the time-distance plot was taken. The upper left panels show light curves extracted from four selected regions (0 - black, 3 - blue, 5 - orange, 7 - red, for ease of presentation), from top to bottom for: Ca~K, H$\alpha$, and Mg, respectively. 
b. (lower panel) Time-distance plot showing motion along the upper and lower flare ribbons for H$\alpha$.   
The velocity derived for the western edge of the main ribbon of $\approx$ 60 km/s is indicated with the red arrow, and a higher velocities of $\approx$300 km/s for smaller expanding regions is indicated with the blue arrow. The derived velocities in the lower ribbon of $\approx$250 km/s is also indicated with the orange arrow.
There is a movie of the HARDCam H$\alpha$  images for Figure 4a (linked in RAA on-line version). The movie starts at 20:55:30 and is shown at a rate of 1 frame per second.
The first and last frames are labeled.
}
  \label{Fig4}
   \end{figure}
  

\subsection{Velocities} 
	We attempted to measure the propagation speed of the erupting ribbons in the photosphere and chromosphere during the flare's evolution in the narrow filter bands (Ca K, Mg and H$\alpha$) using time-distance techniques.  Although it is difficult to distinguish a propagation velocity across the flare ribbon, we find a velocity of $\approx$60 km/s (shown in Figure 4b) for the expansion at the western end of the main flare ribbon in both H$\alpha$ and Ca~K. Similar velocities were found for the Mg band, but were difficult to derive based on its longer cadence, and lower contrast and signal-to-noise. These velocities  compare well with previous studies. \citet{H14} found velocities of filament eruptions observed in H$\alpha$ upto 85 km/s.   More puzzling are smaller features moving as fast as 300 km/s (blue and orange arrows in Fg 4b.).  These large velocities (over 200 km/s) in an expanding ribbon are unusual, and are most likely associated with erupting filaments. Such large velocities of 340 km/s and upto 400 km/s were observed in a filament eruption in the SDO 304, 171 and 193 \AA\ bands by \citet{Z15}.
\section{Discussion}

We investigated the properties of an M3.9 flare observing in multiple narrow and continuum band filters and supplemented with EUV imaging data from SDO. Many issues still remain in our understanding of the details of flare physics (densities, temperatures, and magnetic field configurations) that can only be derived from simultaneous observations at high temporal resolution of many layers of the solar atmosphere.

\subsection{Flare Properties}

We observed intensity increases $\approx$120 to 150\% in the Mg, Ca K and H$\alpha$ narrow band filters during the flare.  Intensity increases for the flare observed in the SDO EUV were several times larger and the GOES X-rays increased a factor of 15 in the soft X-ray band and over a factor of 30 for the harder band. These observed intensity increases in our narrow band filter are slightly lower than those of previous studies \citep{Keys11} and may indicate the majority of the flare emission is at hotter temperatures, as observed by the factor of 4--5 times increase observed with SDO.

   \begin{figure}
   \centering
   \includegraphics[width=12cm, angle=-0]{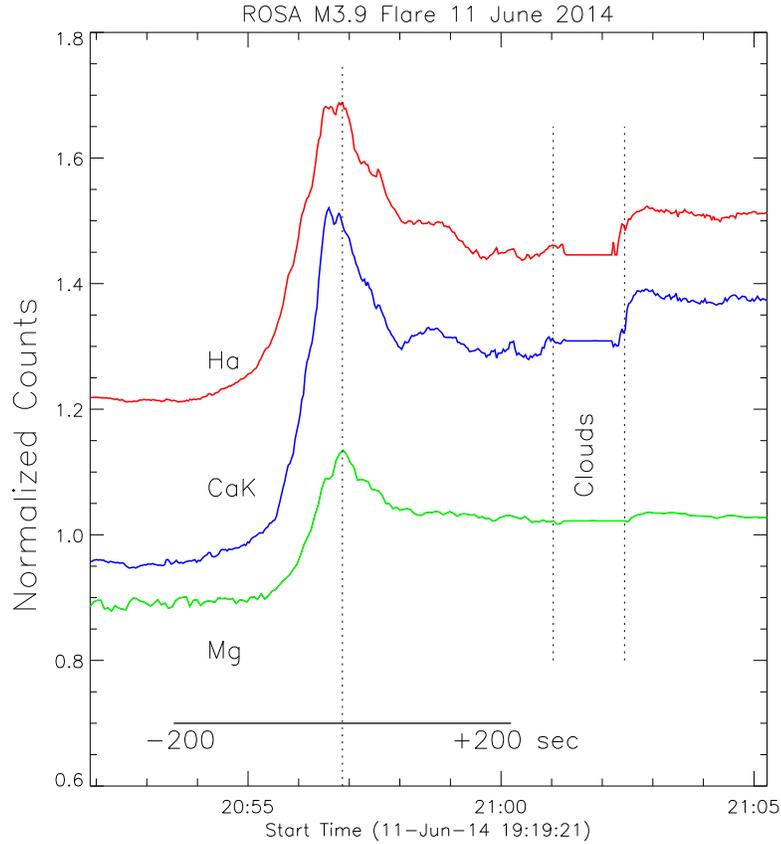}
   \caption{Light curves for different filters for the M3.9 flare. From top to bottom are the H$\alpha$ (red), Ca K (blue), and Mg (green) based plotted in normalized counts (Mg is offset for ease of presentation). No significant delay in the flare peak times is observed between the Mg, Ca K and H$\alpha$ and Mg peaks and is indicated with a vertical dashed line in addition to intervals affected by clouds (see text).  
   }
   \label{Fig5}
   \end{figure}

\subsection{Flare Morphology}
Our observations covered nearly 1 hour before the flare and we are able to observe the detailed morphology of the flare ribbons change on a timescale of seconds.  We classify this as a two-ribbon flare showing 
a lower larger ``main'' ribbon and an upper smaller region observed in our Ca K, Mg, and H$\alpha$ flare emission. Similar flare morpholgy was observed by \citet{BLQ15} using IRIS and SDO observaton of an M-class flare. These ribbons of our flare appear to be related and occur within 30 seconds of each other. We find a length of $\approx$25 Mm for the lower main flare ribbon in H$\alpha$ and Ca K, and Mg bands.  The flare ribbons are very similar to that of white-light flare emitting regions \citep{S76, DB77}  and can be compared in our Mg band to the Mg I b$_2$ observations of \citet{L83}, albeit the 1983 data are at lower spatial resolution.

\subsection{Velocities}
We find almost no difference in the flare's rise times for light curves extracted from regions across the main flare ribbon (Figure 4a) and velocities when comparing various physical regions of the flare using time-distance plots, Also, the rise and decay times between these regions are all very similar (see Figure 4a).  The peaks of the Mg, Ca K and H$\alpha$ emission are all within $\approx$2 seconds (Figure 5).  Typical formation heights for Ca~K and H$\alpha$ are 1300 km and 1500 km, respectively  \citep{VAL81,FAL90}.  This would imply a propagation velocity of about 100 km/s between these 2 layers, but the uncertainties are large. However, it is curious that the Mg emission, which forms near 700 km \citep{S79}; occurs within seconds of the Ca K and H$\alpha$ emission peaks. This implies during the flare this emission could either be being formed at greater heights, similar to the Ca K and H$\alpha$ formation heights. This would be in agreement for flaring models presented by \citet{M90}. This discrepancy may also result from the width of the Mg filter being sensitive to emission at several depths, and we also realize the formation height of chromospheric spectral lines during the flare are significantly different then when comparing to the VAL model for the quiet Sun. 

Similar calculations can be carried out for the SDO data, although we caution that the 12 sec cadence of the AIA EUV channels and the 24 sec cadence of the 1600 \AA\ channel makes the uncertainties in these velocities large, and the SDO filters AIA filters are broadband, capturing emissions from multi-thermal layers with uncertain formation heights. However it gives us a ball-park estimate for propagation velocities to compare with the photosphere and chromosphere. We use \citet{R12} for SDO's formation heights and find velocities consistent with $\sim$100 km/s between emission observed in the SDO 304 \& 1600~\AA\ channels (formation heights for SDO 304 \& 1600 \AA\ are 2200 km and 500 km, respectively), and for propagation between the 304 and SDO 171 ($\Delta$H = 1000km). 
We can also compare our photospheric and chromospheric measurements to those of the transition region and corona as measured by SDO. We can compare the delay between Ca K and the SDO 304 \AA\ band. These 2 layers have a height difference of 900 km for the standard VAL model and this results in a velocity of 75 km/s. If the Ca K formation height is much lower, i.e. 500 km, then the propagation velocity is larger, at $\approx$140 km/s, a slightly higher velocity than a typical M-class flare, and similar to velocities found for other SDO channels. Again we caution the quiet-Sun VAL model is not applicable for the formation heights of chromospheric spectral lines during the flare. The colder SDO channel (SDO 304 with about 50,000 K) had a longer decay time than a hot SDO channel (SDO 193 with about 1 million K), and as we would have expected the SDO 193 flare peaks much later (up to 6 minutes) than SDO 304, since SDO 193 covers the coronal region and SDO 304 the chromosphere. In Figures 6 we illustrate the light curves for several SDO channels along with our ground-based narrow band channels. 

The later emission of the SDO 193 ($\approx1.5\times10^6$ K) and 211 \AA\ channel ($\approx2\times10^6$ K), and the delays observed in the hotter channels (94 and 131 \AA) peak upto 8 minutes later than the cooler SDO channels (e.g. 1600, 304 \AA). These later peaks can be attributed the ''euv late phase'' as described in \citet{W11} and \citet{Liu13a}.  This emission may be caused by different higher, but connected loops that have a different thermal history. Figure 6 also includes the light curve for the 335 \AA\ emission that traces Fe XVI at $\approx$2.5$\times10^6$ for comparison to these works.  Similar delays for the hotter SDO channels were observed by \citet{C15} for a two-ribbon M-class flare, a sample of flares presented by \citet{AS13}, and the X-class flare observed by \citet{Liu14}. We also note that the  GOES bands peaking several minutes later than the narrowband filters and cooler SDO channels is unusual. The GOES 1-8 \AA\ (SXR) light curve is over-plotted in Figure 6 and more closely aligns with the SDO 335, 94 and 131~\AA\ channels and the ''euv late-phase'' emission. Such behavior of the GOES SXR flux lagging emission from the lower solar atmosphere has been observed before for between it and:  the TRACE 1600 \AA\ emission for an M8.0 flare observed by \citet{Liu13b}, SDO 1600 \AA\ emission for a C3.2 flare observed by \citet{Q13}, and with the H$\alpha$ emission for a C4.1 flare observed by \citet{D13}. 
The early impulsive phase of our M3.9 flare observed in narrow band filters (CaK, H$\alpha$, Mg), and cooler SDO bands (304 and 1600 \AA) is readily attributed to rapid heating by non-thermal electron beam precipitation, however the lagging SXR flux is harder to explain. 
\citet{Liu13b} using an 0D EBTEL model, invoke continuously expanding ribbons, emitting in the UV create new flare loops by continuous magnetic reconnection and continue to heat the corona. \citet{Q13} call upon the cooling of hot plasma that has been evaporated into the corona to explain this later phase. The GOES SXR is sensitive to temperatures over 5MK (logT = 6.7) and its later peak agrees well with those observed in the SDO 94, 131 and 335 \AA\ bands, which are sensitive to emission at temperatures greater than 2.5 MK (logT $\approx$ 6.4). Future X-ray solar flare observations need to include high spatial and temporal resolution of the lower solar atmosphere to better model such flare behavior.

   \begin{figure}
   \centering
   \includegraphics[width=12cm, angle=0]{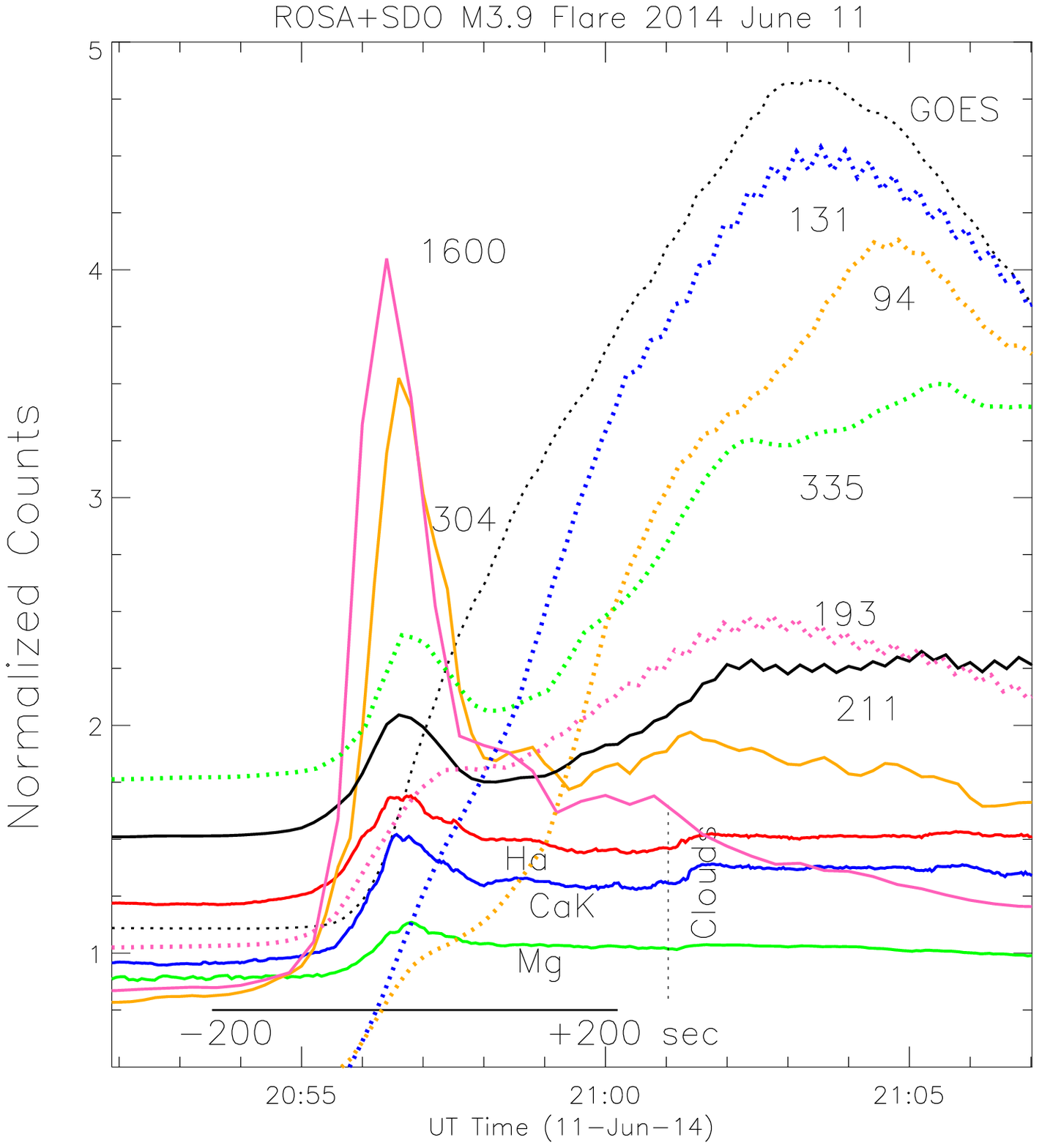}  
   \caption{Shown are the light curves for the cooler (193, 211, 304,  \& 1600 \AA), hotter (94, 131, \& 335 \AA) SDO channels, the GOES soft X-ray band (1--8 \AA; SXR), and our  CaK (blue), Mg (green) and H$\alpha$ (red) narrow-band filter data.      The SDO AIA channels shown are: 304 (orange), 1600 (pink), 211 (black), 193 (pink, dotted line), 335 (green, dotted line), 94 (orange, dotted line), and 131 \AA\ (blue, dotted line).  Individual light curves are normalized by their average counts and are offset for ease of presentation. The re-scaled GOES soft X-ray band (SXR; 1.0 -- 8.0~\AA) is over-plotted (black, dotted line) and aligns with the ''euv late-phase'' emission (see text). 
    }
   \label{Fig6}
   \end{figure}

\subsection{Search for Quasi-periodic pulsations}
We searched for quasi-periodic pulsations (QPP) in our narrow-band filter data using wavelet and Fourier techniques described in \citet{J07}. We found a suggested QPP with periods in the 40--60 second range for the CaK, Mg and  H$\alpha$ bands. 
We present the results of this analysis in Figure 7 for H$\alpha$ (selected from the largest region used in the other analysis for maximum single to noise, although simialr periods are derived for the smaller regions shown in Figure 4a), including the de-trended light curve, wavelet and Fourier power, and confidence level. 
No detections of white light QPP have been found in solar flares and QPP in narrow-band filters have only been suggested. QPPs of about 60 seconds were found after a C-class flare by \citet{Keys11}, and more recently, \citet{KNK15} found quasi-harmonic behavior of QPP observed in an X3.2 flare with intrinsic modes in the 15 to 100 seconds range, and \citet{BLQ15} found periods of $\approx$140 seconds in an M-class flare. The mechanism for QPP are still currently under debate, but involves some form of oscillatory reconnection in the magnetic loops or in MHD modes \citep{MTM12}. More recently, quasi-period slipping was observed during an X-class flare using IRIS and SDO data \citep{LZ15}. The associated periods were found in the 3 to 6 minutes and attributed to tearing mode instability generating oscillatory reconnection that may ultimately be driven by solar p-modes.

   \begin{figure}
   \centering
    \includegraphics[width=12cm, angle=0]{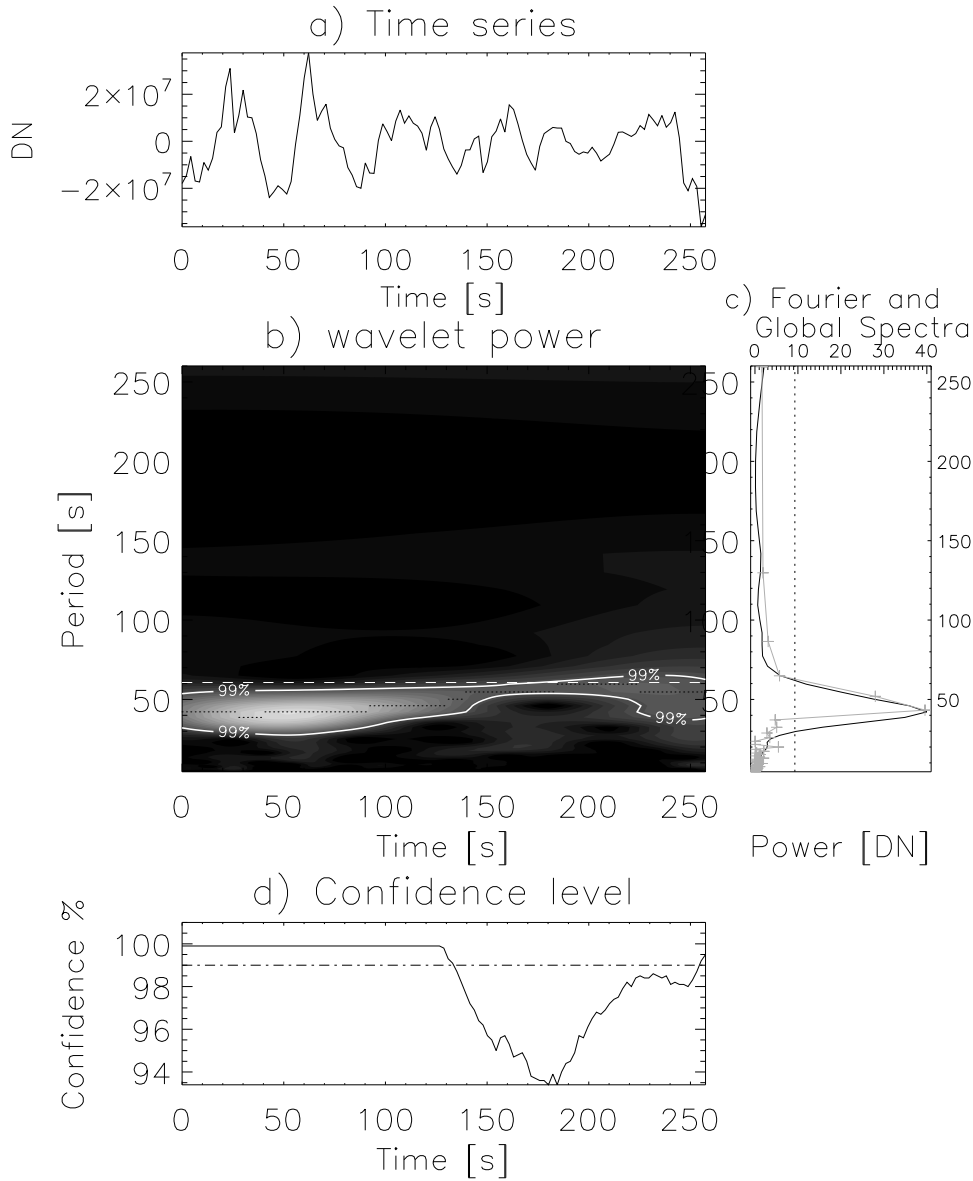}
   \caption{Temporal analysis of the de-trended H$\alpha$ filter light curve. a. The top diagram shows the original H$\alpha$ light curve for $\approx$4 minutes after the flare peak. b. Shown is the wavelet power transform along with locations where detected power is at, or above, the 99\% confidence level are contained within the contours. c) The right-hand-side plot shows the summation of the wavelet power transform over time (full line) and the Fast Fourier power spectrum (crosses) over time, plotted as a function of period. Both methods show marginal detections in the 40$-$60 sec range. The global wavelet (dotted line) and Fourier (dashed dotted line) 95\% significance levels are also plotted. The lowest panel, d. shows the probability levels (1 $-$ p) $\times$ 100 as discussed in Sect. 4.4.}
   \label{Fig7}
   \end{figure}

\section{Summary and Conclusions}
\label{sect:conclusion}

We have presented multi-wavelength observations of the 2014 June 11 M3.9 flare.  We observed intensity increases $\approx$120-150\% in the Mg, Ca K and H$\alpha$ narrow band filters during the flare. Intensity increases for the flare observed in the SDO EUV were several times larger ($\approx$4 times for 304 \AA\ with T $\approx$ 50,000 K),
and the GOES X-rays increased over a factors of 15 and 33 for the soft and hard bands, respectively. The flare morphology observed in these narrow-band filters shows a main flare ribbon and a sympathetic flare ribbon about 7 to 10 Mm away. Only modest delays are found between the onset of emission across the main flare ribbon and upto 30 seconds between the main flare regions and of the nearby sympathetic flare (upper region). The peak flare emission occurs within a few seconds for the Ca K, Mg, and H$\alpha$ bands. Time-distance techniques find propagation velocities of $\approx$60 km/s for the main flare ribbon and velocities as high as 300 km/s in smaller regions of the main flare ribbon that we attribute to filament eruptions. Propagation velocities between the chromosphere and coronal layers, although uncertain, are found to be as large as $\approx$100 km/s. These results and delays and velocities for different coronal heights observed with SDO are consistent with this velocity, and agree well with the simple model of energy propagation versus height, although a more detailed model for the flaring solar atmosphere is needed. Hotter SDO channels show emission with delays of upto 8 minutes and our consistent with the "euv late phase" observed for several M- and X- class flares.  And finally, we detected marginal quasi-periodic pulsations (QPPs) in the 40--60 second range for the Ca~K, Mg and H$\alpha$ bands, and such measurements are currently lacking in the literature, but are important for disentangling the detailed flare-physics. Future high time resolution observations of solar flares (such as those available with DKIST) are important for disentangling the detailed flare-physics.

\begin{acknowledgements}

We thank the NSO Dunn Solar Telescope staff, including Doug Gilliam  for their excellent support for this project, and we dedicate this paper to the late Michael Bradford who's keen eye pointed out this solar flare. 
The work of D.K. was supported by  S\^er Cymru II Part-funded by the European Regional Development Fund through the Welsh Government. 
D.K. acknowledge support from Georgian Shota Rustaveli National Science Foundation project FR17\_323. 
We thank an anonymous referee for suggested improvements to the manuscript. 

\end{acknowledgements}



\label{lastpage}

\end{document}